\documentclass[journal,transmag]{IEEEtran}

\usepackage{hyperref}

\usepackage{cite}
\ifCLASSINFOpdf
   \usepackage[pdftex]{graphicx}
\else
\fi
\usepackage{graphicx}
\usepackage{dcolumn}

\usepackage{amsmath}
\usepackage{array}

\begin{document}
%
\title{Strong Coupling of Magnons to Microwave Photons in Three-Dimensional Printed Resonators}


\author{\IEEEauthorblockN{Vincent Castel\IEEEauthorrefmark{1,2},
Sami Ben Ammar\IEEEauthorrefmark{1},
Alexandre Manchec\IEEEauthorrefmark{3},
Gwendal Cochet\IEEEauthorrefmark{3}, and
Jamal Ben Youssef\IEEEauthorrefmark{4,2}}
\IEEEauthorblockA{\IEEEauthorrefmark{1}IMT Atlantique, Technopole Brest-Iroise, CS 83818, 29238 Brest Cedex 3, France}
\IEEEauthorblockA{\IEEEauthorrefmark{2}Lab-STICC (UMR 6285), CNRS, Technopole Brest-Iroise, CS 83818, 29238 Brest Cedex 3, France}
\IEEEauthorblockA{\IEEEauthorrefmark{3}Elliptika (GTID), 29200 Brest, France}
\IEEEauthorblockA{\IEEEauthorrefmark{4}Universit\'e de Bretagne occidentale, 6 Avenue Le Gorgeu, 29200 Brest}
\thanks{Corresponding author: V. Castel (email: vincent.castel@imt-atlantique.fr)}}

%



\IEEEtitleabstractindextext{%
\begin{abstract}
We report on ferromagnetic resonant mode hybridization in re-entrant cavities made with a commercial three-dimensional (3D) printer, followed by conventional 3D metalization with copper and tin. The cavity volume was only 7$\%$ that of a standard cavity resonating at the same frequency, while maintaining a high quality factor. Simulations were in very good agreement. We obtained an effective coupling of about 40 MHz in two cavities at room temperature. These experimental results demonstrate the utility of tunable filters based on complex 3D printed cavities.
\end{abstract}

}

\maketitle

\IEEEdisplaynontitleabstractindextext

%
\IEEEpeerreviewmaketitle

\section{Introduction}

\IEEEPARstart{I}{}n recent years, a new light-matter interaction platform has been developed\cite{Tabuchi2014,Bai2015,Maier-Flaig2016,Bai2016,Cao2014} by combining microwave photons residing in a resonant cavity with magnons (viewed as a quantized spin wave) in magnetic materials with a high spin density like Yttrium Iron Garnet (YIG). Here, the strong coupling (even ultrastrong\cite{refbig,PhysRevLett.111.127003}) between magnons with photons results in a hybrid light-matter quasiparticle: magnon-polariton. The use of a standard 3D cavity (developed in this platform) for microwave applications seems difficult to combine with concepts of integration, compactness, cost and miniaturization. Here, we propose to take advantage of the geometric configuration of re-entrant microwave cavity to counterbalance this problem, which is becoming more and more critical by reducing the working frequency. Microwave re-entrant cavities offer many advantages in terms of frequency tunability, volume reduction, control of the microwave magnetic field distribution, while keeping a high Q factor. Such cavities have been used for decades for a variety of applications like high Q resonator or sensors, and in different research area from the observation of spin wave propagation\cite{Eshbach1962} to studies of cavity magnon-polaritons\cite{PhysRevB.97.184420,PhysRevB.97.155129,PhysRevApplied.2.054002}.

The main purpose of this work is to demonstrate the strong coupling signature between a YIG thin film resonator and photons generated by a 3D re-entrant cavity elaborated by additive manufacturing.

\section{3D printed microwave re-entrant cavity}
\begin{figure}
	\includegraphics[width=9cm]{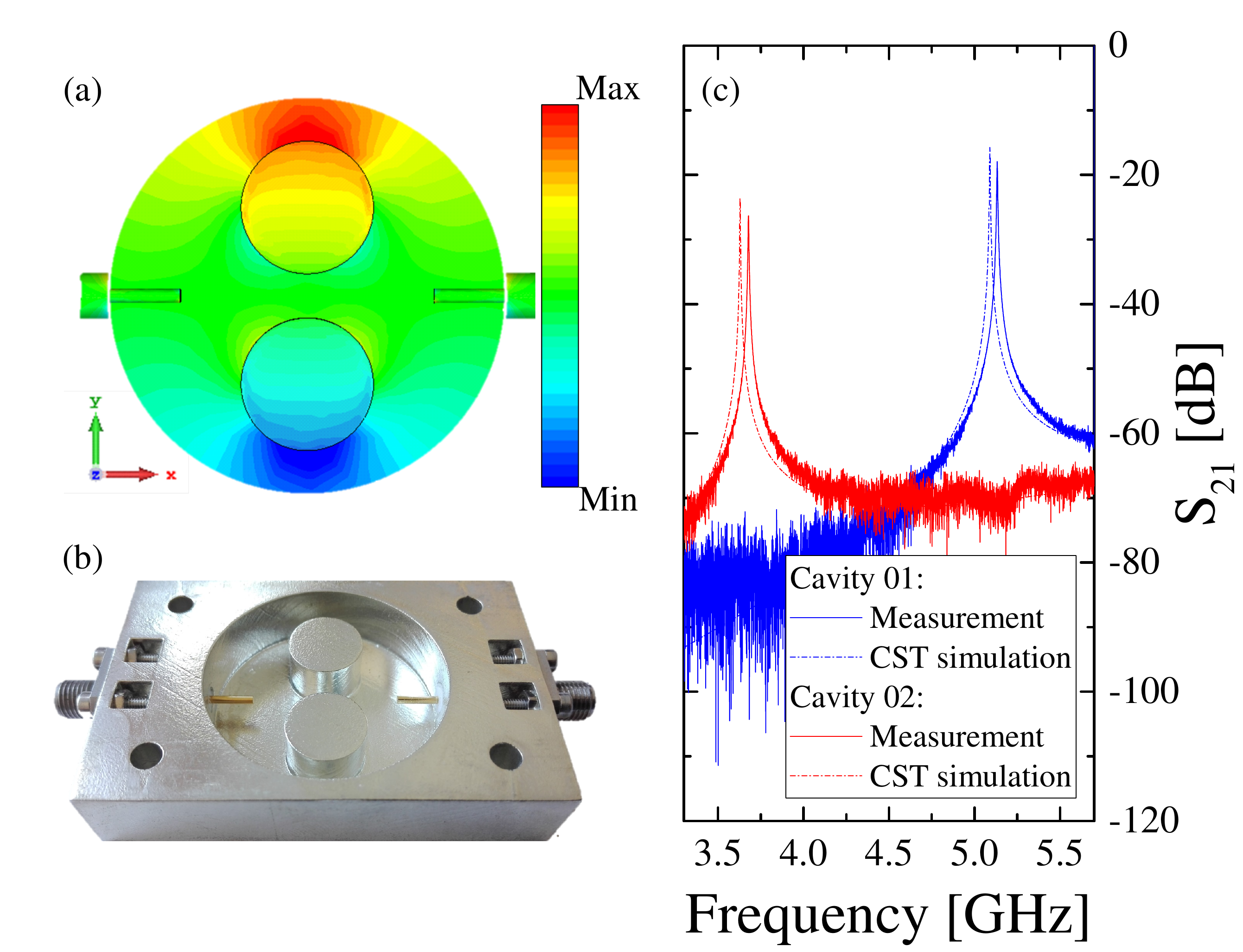}
	\caption{\label{fig:Fig1} 
		(a) Simulated microwave (MW) magnetic field distribution (X component) of the dark mode at the resonant frequency of cavity 01. (b) Photograph of the 3D printed cavity (without lid) including straight radio-frequency (RF) connector. (c) Frequency dependence of the $\textrm{S}_{21}$ of cavity 01 (blue) and 02 (red). Solid and dash lines correspond respectively to the measurement at room temperature and CST simulation.   
	}
\end{figure}

The use of posts inside a 3D cavity permits to reduce its final volume and to offer the possibility to control precisely the working frequency by keeping a high Q factor. In such configuration, most of the microwave (MW) electric field is concentrated in the gap (creates between the lid and the top of inner posts), while all of the MW magnetic field is distributed around the post, with a rapid decrease in amplitude, as shown in Fig.\ref{fig:Fig1} (a). The height of our first 3D printed cavity prototype is based on standard commercial radio-frequency (RF) connector dimension features in order to avoid complications in the microwave excitation and to ensure a very good reproducibility of S parameter measurements. In other words, the present cavity is not the minimal volume that we could achieve by using 2 inner posts but it offers already a volume reduction by one order from 2.29 10$^{-4}$ m$^3$ (without posts) to 1.018 10$^{-5}$ m$^3$ (with 2 posts), working at the same frequency. The radius of the cavities ($R_{cavity}$) and the posts ($R_{post}$) are 18 and 6 mm, respectively. The inner height for both cavities is $h$=10 mm and posts positions are identical. The difference between cavities (refereed as cavity 01 and 02) is only associated to the gap width $d$ between the lid and posts which directly impacts the working frequency ($d$ is equal to 5.0 and 2.07 mm for cavity 01 and 02, respectively). A commercial solver of Maxwell's equations (CST microwave studio) has been used to design cavities and we estimate that the resonant frequency of the first mode could be tune from 2 to 5.5 GHz by changing $d$ from 0.5 to 6 mm. The large frequency tuning capability (which could be optimized) makes this 3D printed cavity very attractive for telecommunication systems and competitive in comparison with standard reconfigurable solution such as lump tuning\cite{7539960,6656009} components or piezoelectric transducers\cite{4264292}. The 3D printed cavities used in this work must be modified to develop a reconfigurable function. This functionality can be achieved by integrating microwave tuning elements that will serve as adjustable inner posts.

Our 3D cavities have been elaborated by a standard stereolithography 3D printer (Formlabs Form 2). Each of them received two final steps which consist to clean residual resin in a IPA alcohol bath and to treat them in a UV chamber including a heat treatment. Conventional 3D metallization procedure of plastic elements has been adapted to our goal in order to produce a full metallization of plastic re-entrant cavities described as follows: (i) Dry etching by sandblasting (enhancement of the roughness), (ii) Surface activation with Pd$^{2+}$ solution, (iii) Autocatalytic bath of copper: Pd particles act as catalyst sites and permit the growth of an homogeneous layer of Cu which spontaneously end up when reaching a 3 $\mu$m thickness, (iv) Standard electrodeposition process of Cu (5-10 $\mu$m) including a Tin finish of 8 $\mu$m to ensure a high electrical conductivity. The final metallic layer deposit on each cavity consists of a thickness of 16 to 21 $\mu$m which correspond to at least seven skin-depths at 1 GHz\cite{8371426}. Fig.\ref{fig:Fig1} (b) shows a picture of cavity 02 without the lid on top. Note that for more complex 3D geometries, dry etching by sandblasting (i) is replaced by a chemical bath of Chromium. The pin length of the RF connector is based on a compromise between Q factor, homogeneity of the MW magnetic field distribution at the resonant condition, and the strength of the transmitted signal. Such a cavity is characterized by the spatial separation of the MW electric and magnetic field distribution. The inner post acts as an individual microwave resonator with its own characteristics (inductance and capacitance). Therefore, the entire system has two modes: a dark mode and a bright mode. In the bright mode, the MW magnetic field generated by both posts constructively interferes and enhances the total magnetic field between the posts. The RF excitation configuration that we have used in the present work (straight connectors) allows us to observe only the dark mode of the cavity which is based on the destructive interference of MW magnetic fields. The lack of a MW magnetic field in the center of the cavity is well represented by Fig.\ref{fig:Fig1} (a) showing the simulated distribution of this field at the resonant frequency. Note that the bright mode of our cavities can only be detected by changing at least one connector to a loop probe. 

The measurement at room temperature of the frequency dependence of the $\textrm{S}_{21}$ of cavity 01 (blue) and 02 (red) is represented in Fig.\ref{fig:Fig1} (c). A good agreement is found with simulations (dashed lines) for which we approximated the cavity material as a bulk Tin object. The mismatch in terms of resonant frequency between measurement and simulation observed in Fig.\ref{fig:Fig1} (c) and highlighted in Table \ref{table1} is due to the deviation of the dimension of the 3D printed cavities from the simulation model (used for the printer). A perfect match can be found (not shown) by taking into account an overall enhancement in the structure of 140 $\mu$m, which corresponds to the size of the 3D printer's laser spot. The reduction of the Q factor is mainly due to high roughness of the coating metal (about 12.5 $\mu$m) which presents an electrical conductivity of 7.88 $10^6$ S/m. In order to adjust the conductor loss, we estimated a correction factor Cs\cite{rugo} which takes into account the surface roughness and the skin depth. CST simulation has been performed by adjusting the electrical conductivity at 4 $10^6$ S/m (measured value divided by Cs=1.97) and permit to reproduce the observed Q factor.
 
\begin{table}[ht]
	\caption{3D printed cavities specifications}
	\label{table1}
	\centering
	\begin{tabular}{c|c|c||c|c}	
		
		& \multicolumn{2}{c||}{Cavity 01} &\multicolumn{2}{c}{Cavity 02}  \\
		& Meas. & CST   & Meas. & CST \\
		\hline
		& & & &  \\ [-6.5pt]
	   Bright mode $F_{0}$ [GHz]        &-    &6.642 & -      & 4.867\\
	   		\hline
	   			& & & &  \\ [-6.5pt]
	   Dark mode $F_{0}$ [GHz]         &5.133&5.090 & 3.677  &3.628\\ 
		$\textrm{S}_{21}$ @ $F_{0}$ [dB]         &-17.92 & -15.64 & -26.17 &-23.65\\
		$\Delta F_{-3dB}$ [dB]          &4.56 & 3.30 & 4.43 &2.90  \\
		Q                               &1125.6 &1542 &830 &1251 \\
		$\beta$ [10$ ^{-4} $]           &4.442 &3.242 &6.024 & 3.997 \\

	\end{tabular}
\end{table}

\section{Experimental details}
The use of ferrimagnetic insulating material consists of a single-crystal (111) Y$_3$Fe$_5$O$_{12}$ (YIG) film of 9 $\mu$m grown on a (111) Gd$_3$Ga$_5$O$_{12}$ (GGG) substrate by liquid phase epitaxy (LPE). The sample has been cut into a rectangular shape (4 mm$\times$6 mm) by using a Nd-YAG laser working at 8 W. YIG exhibits spin waves with the highest quality factors among all magnetic materials. It is also free from dissipation due to conduction electrons, resulting in a narrow linewidth. The long lifetime of collective spin excitation in YIG has attracted a lot of interest not only for the field of spintronics but also for magnon engineering in the quantum regime.
\begin{figure}[ht]
	\includegraphics[width=9cm]{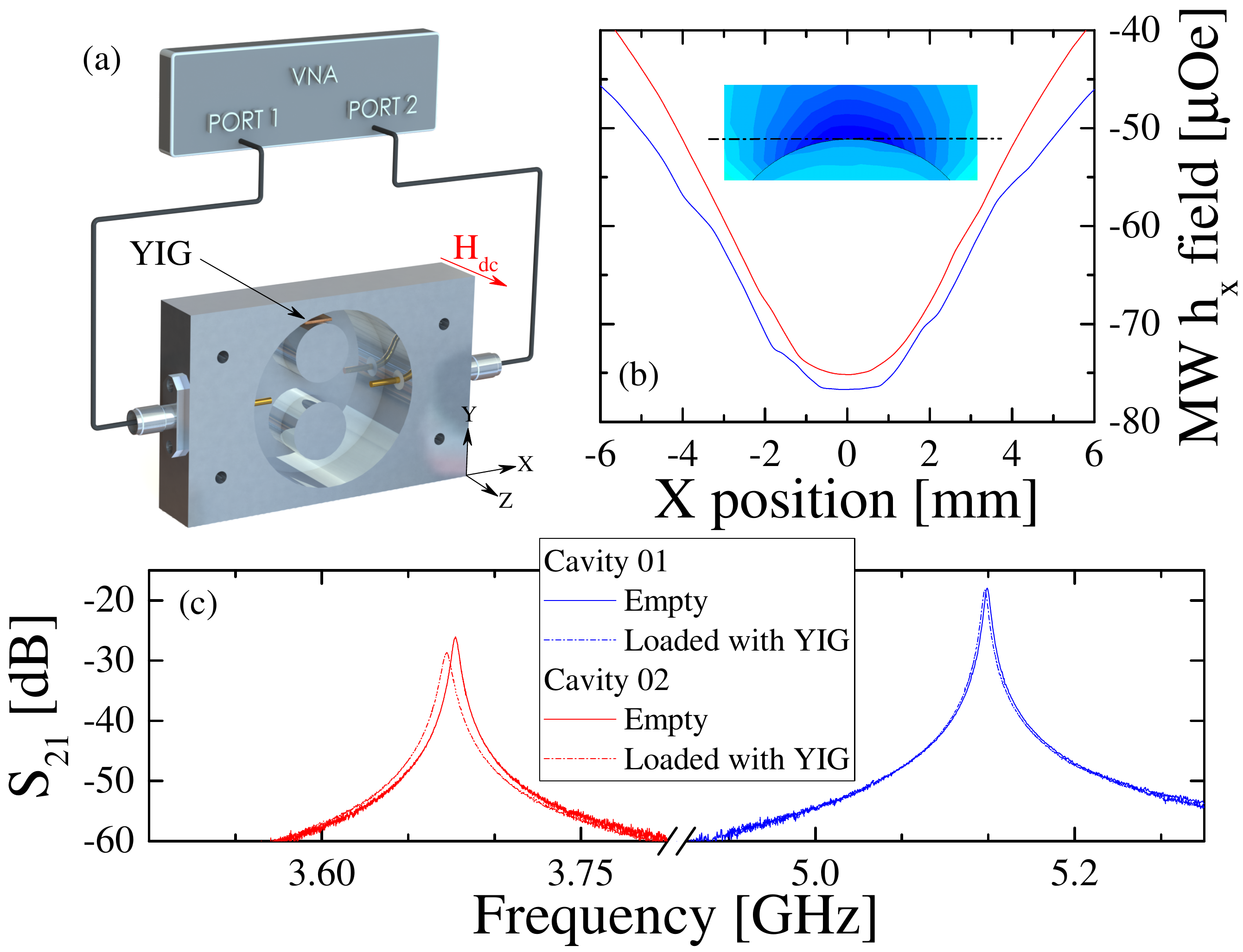}
	\caption{\label{fig:Fig2} 
		(a) Measurement schematic. The YIG sample is placed at the bottom of the cavity. Its surface is parallel to the plan (X,Z) and on top of the upper post. This position permits to keep a static magnetic field ($H_{dc}$) perpendicular to the MW magnetic field. A calibrated VNA permit to extract S parameters. (b) Microwave magnetic field profile on X axis taking from simulation shown in Fig.\ref{fig:Fig1} (a). The profile section corresponds to the value of the field taken at a height of 2 mm (middle of the sample) from the bottom of both cavities and parallel to the X axis. The profile section is indicated by the dashed line. (c) Measurement at room temperature of the frequency dependence of $\textrm{S}_{21}$ of both cavities, with and without YIG sample in it.}
\end{figure}
The plane of the YIG film is placed between the cavity wall and the post, as shown in Fig.\ref{fig:Fig2} (a), where a maximum of MW magnetic field is generated (Fig.\ref{fig:Fig1} (a)). The crystallographic axis [1,1,$\bar{2}$] is parallel to the X axis and the static magnetic field, $H_{dc}$, is applied along Z, the normal of the cavity. By using this configuration, we satisfy the FMR condition which is described by the precession frequency of the uniform mode in an in-plane magnetized ferromagnetic film defines as $F_{r}=(\gamma/2\pi)\sqrt{H(H+4\pi M_{s})}$, where $4\pi M_{s}$ and $\gamma$ are respectively the saturation magnetization and the gyromagnetic ratio. In Fig.\ref{fig:Fig2} (a), PORT 1 and PORT 2 illustrate the two ports of the vector network analyzer (VNA) that were calibrated, including the two cables. The frequency range is set from 3 to 6 GHz at a microwave power of $P=-10$ dBm and $P=-5.5$ dBm for cavity 01 and 02, respectively. The input power was enhanced from cavity 01 to 02 in order to excite the FMR condition with a similar MW magnetic field magnitude. The appropriate input power was determined via CST simulation. Fig.\ref{fig:Fig2} (b) represents the profile section of the MW $h_X$ (at the selected input power) along X axis seen by the YIG film and taken at a height of 2 mm from the bottom of both cavities. It is clear that the MW magnetic field distribution of the dark mode is less homogeneous than the bright mode (observed in the middle of the cavity) with an evolution of MW $h_X$ about 10 $\mu$Oe (13 $\%$) from the middle to the edge of the YIG sample. Note that the distribution of this component along the Z axis (along the YIG height) does not present a similar behavior as function of the post height. From the top to the bottom of the YIG sample, the MW magnetic field is affected by a reduction of 17 $\%$ for cavity 01 whereas for cavity 02 (gap d of 2.07 mm) the reduction is only of 5 $\%$. Nevertheless, the reproducibility, the mode excitation features, and the use of RF standard connector facilitate development of RF applications. The bright mode is very attractive but the loop probe must be well designed in order to be able to detect this mode without affecting the frequency response with an additional resonance associated with the length of the loop. 

Before studying the coupling strength between magnons and photons, we characterized the updated features of the two spin waves/cavities systems in the absence of a static magnetic field. Fig.\ref{fig:Fig2} (c) represents the measurement at room temperature of the frequency dependence of $\textrm{S}_{21}$ for both cavities, with (loaded) and without (empty) YIG sample in it. A minor impact in the frequency response was observed for the cavity 01, which is not the case for cavity 02 (resonating at lower frequency). The frequency behaviour of cavity 02 loaded with the YIG film (exactly the same sample used in cavity 01) is not clearly explained. Further simulations have been done by including a YIG film (with the GGG substrate as well). The position of the sample defines by the height respect to the cavity bottom and the angle arbitrarily created with X axis induces a variation of cavities features. Nevertheless, simulated variations respect to the initial position of the sample are very small compared to the experimental findings summarized in Table \ref{table2}. The assumptions used to create the 3D model of the cavity for the simulations appear to be correct with respect to the results obtained for cavity 01 which resonates at high frequency. In order to understand the deviation observed in cavity 02, feather investigations are required to determine the frequency dependence of the microwave response with and without YIG by using a unique cavity. The flexibility offered by 3D printing will allow us to develop a cavity that will include tunable post height elements.
\begin{table}[ht]
	\caption{Strong coupling features}
	\label{table2}
	\centering
	\begin{tabular}{c|c|c||c|c}	
	& \multicolumn{2}{c||}{Cavity 01} &\multicolumn{2}{c}{Cavity 02}  \\
		& Empty & Loaded   & Empty & Loaded \\
		\hline
		& & & &  \\ [-6.5pt]
		$F_{0}$ [GHz]                  &5.133  &5.130   & 3.677  &3.672\\ 
		$\textrm{S}_{21}$ @ $F_{0}$ [dB]        &-17.92 &-18.32  & -26.17 &-28.67\\
		$\Delta F_{-3dB}$ [dB]         &4.56   & 5.00   & 4.43   &6.00  \\
		Q                              &1125.6 &1026    &830     &612 \\
		$\beta$ [10$ ^{-4} $]          &4.442  &4.873   &6.024   & 8.17 \\
		$k$                            &-      &0.125   &-       & 0.156 \\
		$\frac{g_{eff}}{2\pi}$ [MHz]   &-      &40.10   &-       & 44.75 \\
	\end{tabular}
\end{table}

\section{Strong coupling regime}

\begin{figure}
	\includegraphics[width=9cm]{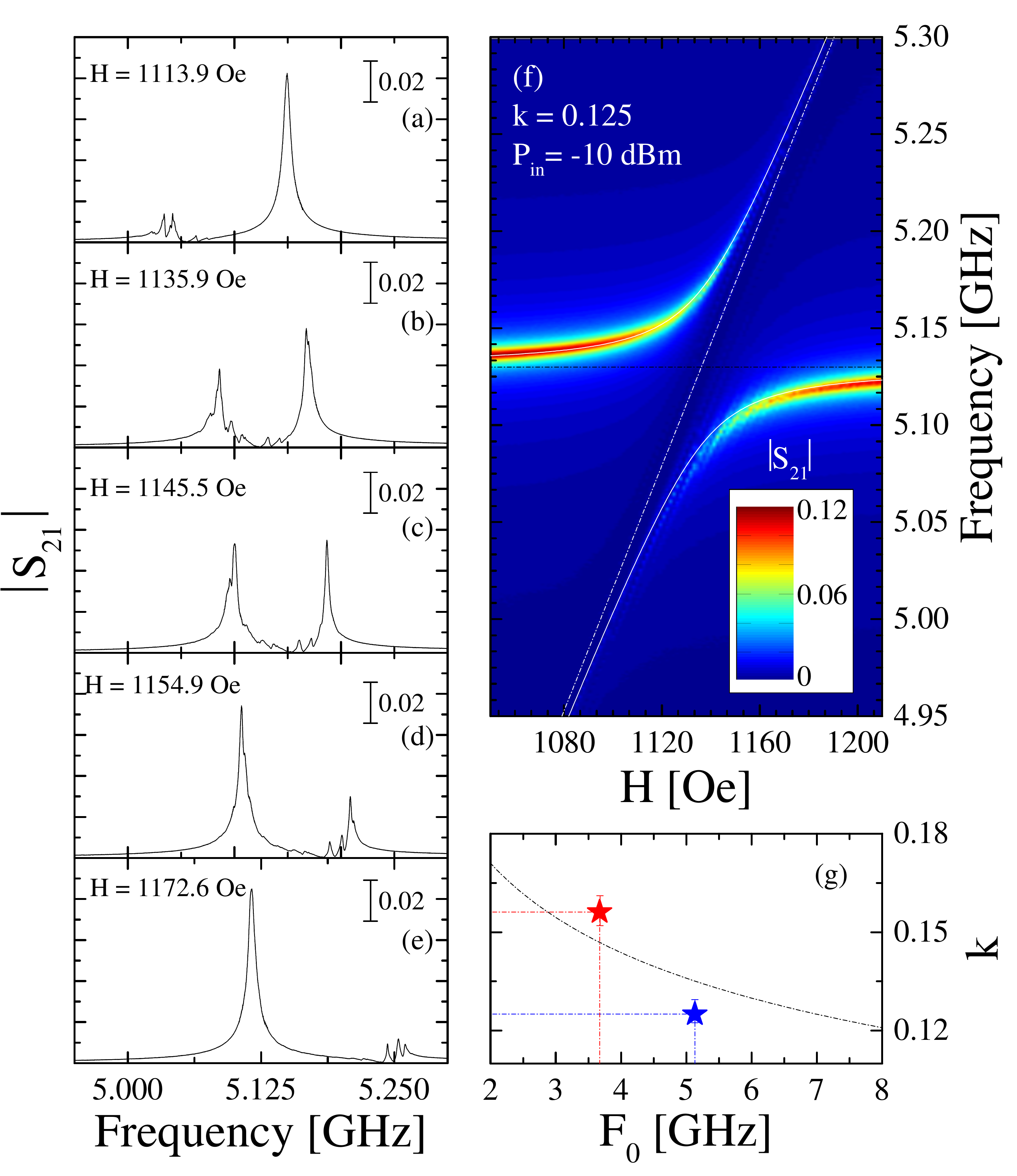}
	\caption{\label{fig:Fig3} 
		(a)-(e) Measurement of the frequency dependence of the $|\textrm{S}_{21}|$ of the assembled YIG-Cavity 01 resonator system for different values of the static magnetic field applied along the posts, above and below the resonant magnetic field, $H_{res}$. (f) Experimental observation of the anti-crossing region at room temperature: Measurement of the transmission amplitude $|\textrm{S}_{21}|$ (defined by the colour bar) of the resonator system as function of the frequency and the applied static magnetic field. (g) Dash-dot line corresponds to the frequency dependence of the coupling strength, $k$, based on Eq.\ref{kcal}. Blue and red stars are associated to the experimental determination of $k$ of Cavity 01 and 02, respectively.
	}
\end{figure}

In order to evaluate the coupling strength between magnons and photons generated by 3D printed cavities, we investigated the frequency response of the system as a function of the applied magnetic field. Figure \ref{fig:Fig3} (a) to (e) show measurements of the frequency dependence of the $|\textrm{S}_{21}|$ of the YIG/Cavity 01 system for different values of the static magnetic field applied along the posts, above and below the resonant magnetic field. The interaction between magnons and photons led to the following features: (i) hybridization of resonances, (ii) annihilation of the resonance at $F_{0}$, and (iii) generation of two resonances at $F_{1}$ and $F_{2}$. At the resonant condition, the frequency gap, $F_{gap}$, between $F_{1}$ and $F_{2}$ is directly linked to the coupling strength of the system ($F_{gap}=\dfrac{g_{eff}}{\pi}$). Based on the harmonic coupling model \cite{Harder2016}, we can define the upper ($F_{1}$) and lower ($F_{2}$) branches by $F_{1,2}=\dfrac{1}{2}\left[ \left( F_{0}+F_{r}\right) \pm \sqrt{\left( F_{0}-F_{r}\right)^{2}+k^{4}F_{0}^{2}}\right]$. Hybridization resonance modes feature observed in Fig. \ref{fig:Fig3} (a) to (e) confirmed the strong coupling regime between both resonators. It is clear that the effective coupling, $\dfrac{g_{eff}}{2\pi}$, is greater than losses of both resonators. It should be noted that the response of the YIG resonator does not reflect the intrinsic magnetic loss but includes the inhomogeneous broadening due to the excitation of several modes. Even so, the average linewidth for the magnetic resonator is smaller than $\dfrac{g_{eff}}{2\pi}$. Similar behaviour has been observed with the YIG/Cavity 02.

A fine magnetic field step was used to demonstrate a clear signature of the strong coupling regime. Figure \ref{fig:Fig3} (f), which represents the experimental frequency dependence of $|\textrm{S}_{21}|$ with respect to the applied magnetic field measured at room temperature, demonstrates the strong coupling regime via the anti-crossing fingerprint. The parameter $k$ corresponds to the coupling strength which is linked to the experimental data $\dfrac{g_{eff}}{2\pi}$ by the following equation \cite{Harder2016}: $F_{gap}=F_{2}-F_{1}=k^{2}F_{0}$. A good agreement of $F_{1,2}$ (solid white lines) with experimental data is obtained with $k$=0.125 (cavity 01) and 0.156 (cavity 02), the saturation magnetization $4\pi M_{s}$=1775 G, and the gyromagnetic ratio $\gamma$=1.8 10$^7$ rad Oe$^{-1}$s$^{-1}$. The color plot is associated with $|\textrm{S}_{21}|$ (linear) for which the red area corresponds to a transmission amplitude of 12$\%$. Features of YIG/cavity systems are summarized in Table \ref{table2} and also includes cavities losses $\beta$.

The purpose of this work is not to follow the different excited spin waves mode (and to extract in the meantime the associated wave vectors) but to compare the experimental determination of the coupling strength, $k$, as function of the resonant frequency to the following equation\cite{Maier-Flaig2016,Tabuchi2014}:
\begin{equation}
\frac{g_{eff}}{2\pi}=\frac{\eta}{4\pi}\gamma\sqrt{\frac{\hbar\omega_{0}\mu_{0}}{V_{c}}}\sqrt{2Ns},
\label{kcal} 
\end{equation}
where $\mu_{0}$ is the permeability of the vacuum, $V_{c}$ corresponds to the volume of the cavity, $N$ is the total number of spins (linked to the excited YIG volume), and $s$=(5/2) is the spin number of the ground state Fe$^{3+}$ ion in YIG. The coefficient $\eta\leq 1$ describes the spatial overlap and polarization matching conditions between the microwave field and the magnon mode. In a previous work, we demonstrated an exceptional point in a notch filter coupled to a YIG/Platinum system by thermally control the magnetic damping (magnetic losses by analogy with $\beta$) with a current-induced heating method\cite{PhysRevB.96.064407}. The effective strength found in this latter work were higher compared to the present work. It is due to the microwave configuration of the planar resonator for which an effective small volume can be estimated for a dimensional transmission-line cavity\cite{Schoelkopf2008} with $V_{c}=\pi r^{2}\lambda/2$. Equation \ref{kcal} shows that it is possible to adjust the effective coupling with the number of spin (volume of YIG) \cite{Cao2014,NotchVincent}, the volume of cavity \cite{Schoelkopf2008}, or the resonant frequency. Figure \ref{fig:Fig3} (g) represents the experimental frequency dependence of $k$ based of both YIG/cavities systems which present a similar $V_{c}$ and the same number of spins. It shows that $k$ decrease with the enhancement of the frequency. This latter dependence can be understood by using Eq.\ref{kcal} which is represented in the figure with the dash-dot line. The next generation of 3D printed re-entrant cavity based on our processing protocol will include adjustable post height elements to monitor the frequency dependence of $k$ over a wide range with more accuracy (single cavity).

\section{Conclusion}
We reported mode hybridization between magnons and photons generated by a full metallized 3D printed re-entrant cavities, elaborated with a commercial stereolithography 3D printer. A frequency dependence of the coupling strength was observed with the two YIG/Cavity systems which differ only with the inner post height (similar volume). 3D additive manufacturing opens an innovative path for the development of a new class of 3D tunable filters by combining the advantages of re-entrant cavities. Such elaboration procedure will also improve the integration of YIG/Pt system without drastically increasing the intrinsic loss rate of the cavity.

\section*{Acknowledgments}
This work is part of the research program supported by the European Union through the European Regional Development Fund (ERDF), by Ministry of Higher Education and Research, Brittany and Rennes M\'etropole, through the CPER Project SOPHIE/STIC $\&$ Ondes, and by Grant-in-Aid for Scientific Research of the JSPS (Grant Nos. 25247056, 25220910, and 26103006).

\bibliographystyle{IEEEtran}
\bibliography{SpinCavity}


%

%
%
%
%
%

\ifCLASSOPTIONcaptionsoff
  \newpage
\fi

\end{document}